\documentclass[pra,aps,twocolumn]{revtex4}
\usepackage{graphicx}
\usepackage{amsmath}
\usepackage{amsfonts}
\usepackage{amssymb}
\usepackage{color}

\def\M{\mathsf{M}}
\def\A{\mathsf{A}}
\def\B{\mathsf{B}}
\def\C{\mathsf{C}}
\def\D{\mathsf{D}}
\def\I{\mathsf{I}}
\def\O{\mathsf{O}}

\def\AD{\mathsf{AD}}
\def\CB{\mathsf{CB}}

\def\a{\mathsf{a}}
\def\b{\mathsf{b}}
\def\ab{\mathsf{a,b}}
\def\dd{\mathrm{d}}

\def\inp{\mathrm{in}}
\def\out{\mathrm{out}}
\def\at{\mathrm{at}}
\def\sep{\mathrm{sep}}
\def\li{\mathrm{li}}
\def\eff{\mathrm{eff}}

\begin{document}

\title{Phases and relativity in atomic gravimetry}
\author{M.-T. Jaekel}
\affiliation{Laboratoire de Physique Th\'{e}orique de l'ENS, UPMC, CNRS,
24 rue Lhomond, F-75231 Paris, France}
\author{B. Lamine}
\affiliation{Laboratoire Kastler-Brossel, CNRS, ENS, UPMC, Campus
Jussieu, F-75252 Paris, France}
\author{S. Reynaud}
\affiliation{Laboratoire Kastler-Brossel, CNRS, ENS, UPMC, Campus
Jussieu, F-75252 Paris, France}

\begin{abstract}
{The phase observable measured by an atomic gravimeter built up on
stimulated Raman transitions is discussed in a fully relativistic
context. It is written in terms of laser phases which are invariant
under relativistic gauge transformations. The dephasing $\Phi$ is
the sum of light and atomic contributions which are connected to one
another through their interplay with conservation laws at the
interaction vertices. In the case of a closed geometry, a compact
form of the dephasing is written in terms of a Legendre transform of
the laser phases. These general expressions are illustrated by
discussing two techniques used for compensating the Doppler shift,
one corresponding to chirped frequencies and the other one to ramped
variations.}
\end{abstract}

\maketitle

\section{Introduction}

After years of development, matter-wave interferometers have become
accurate tools for measuring gravity fields. Matter-wave gravimeters
were realized by using first neutron matter-waves
\cite{Colella75,Staudenmann80,Klein83,Greenberger83} and then cold
atoms matter-waves \cite{Kasevich91,Fixler07,Bouyer07,LeGouet0710}.
These instruments are expected to open new roads to fundamental
tests in relativistic, gravitational or quantum physics
\cite{Dimopoulos07,Arvanitaki08,Ertmer09,Wolf09,Amelino09,Stockton11}.
They have been studied in a number of detailed publications
\cite{Borde89,Kasevich92,CohenTannoudji93,Storey94,Wolf99,%
Peters01,Borde02,Borde08,Dimopoulos08,Cronin09}.

Atomic interferometry experiments may be used to test the
universality of free fall (UFF) by comparing the free fall of
quantum bodies to that of classical test masses \cite{Peters99}. A
recent claim \cite{Mueller10} has however blurred this up to then
clear interpretation. The claim changes the interpretation of atom
gravimetry experiments now considered as measuring the redshift on
the quantum clock operating at a high Compton frequency. After this
re-interpretation, these experiments would be considered as testing
the universality of clock rates (UCR). The new interpretation has
been contradicted \cite{WolfMueller10} and opened a debate on the
significance of atom interferometry measurements in the
context of relativity \cite{Sinha11,Giulini11,Unni11,Wolf11,%
HohenseeWolf12,Hohensee11,Wolf12,Schleich12}.

It is the aim of the present paper to present calculations of phase
observables in a relativistic context and to stress points which are
important for any study of this problem in metric theories. First,
the phase observables involved in atomic interferometry are
\emph{gauge invariant} quantities. We will of course make references
to coordinate systems but an important aim will be to write
equations or statements independently of the specific choice of the
coordinate system. Then, the expressions of the phase observables is
intimately related to the conservation laws which constrain the
description of interaction vertices. We will thus be careful to give
a full account of conservation laws, and to make explicit their
intimate connections with the expressions of phase observables.

The main focus of this paper is the effect of a uniform acceleration
and we will in particular disregard the effect of gravity gradients,
gravitational waves or rotations. This means that we study the
unperturbed operation of atomic interferometers designed to measure
the perturbation associated with a Riemann curvature
\cite{Lamine02,Delva06,Dimopoulos08b,Tino11,MIGA}.

\section{Relativistic gravimetry}

We consider an atomic gravimeter based on stimulated Raman
interactions \cite{Kasevich92} which monitors the free fall of atoms
with respect to an experimental platform at rest in the laboratory
frame $\mathcal{R}$. We focus the attention on the case of a
gravimeter measuring a stationary acceleration with no rotation and
no curvature. This case may be represented by the M{\o}ller metric
\cite{Moller52,Rohrlich63}
\begin{equation}
\dd s^{2}=\left( 1+\gamma z \right) ^{2}\left( c\dd t\right)
^{2}-\left( \dd x^{2}+\dd y^{2}+\dd z^{2}\right) \,.
\label{metricLab}
\end{equation}%
The metric element $\dd s^2$ defines proper time \cite{Landau}, with
$\dd s^2=c^2\dd \tau^2$. The parameter $\gamma$ has the meaning of
an acceleration divided by $c^2$.

As the Riemann curvature vanishes, the same physical situation may
be described in a Minkowski frame $\overline{\mathcal{R}}$
\begin{equation}
\dd s^{2}=\left( c\dd \overline{t}\right) ^{2}-\left( \dd
\overline{x}^{2}+\dd \overline{y}^{2}+\dd \overline{z}
^{2}\right)\,. \label{metricBar}
\end{equation}%
The M{\o}ller map from \eqref{metricLab} to \eqref{metricBar} may be
written under the reciprocal forms ($x,y$ unaffected)
\begin{eqnarray}
\label{CoorTra} && 1-\eta\gamma\overline{u}^\eta=\left(1+\gamma
z\right) e^{-\eta\gamma ct} \; , \\
&& 2 \gamma ct = \ln\left(1+\gamma\overline{u}^-\right)
-\ln\left(1-\gamma\overline{u}^+\right) \; , \notag \\
\label{CoorTraInv} && 1+\gamma z=\sqrt{\left(
1-\gamma\overline{u}^+\right) \left( 1+\gamma\overline{u}^-\right) }
\; , \notag
\end{eqnarray}%
where we have introduced the two light cone variables
\begin{equation}
\overline{u}^\eta \equiv c\overline{t}-\eta \overline{z} \; ,\quad
\eta =\pm 1 \,.
\end{equation}
$\mathcal{R}$ is the laboratory frame in which the experimental
platform is at rest while $\overline{\mathcal{R}}$ is the free fall
frame in which atomic probes have inertial motions. They are
convenient for different parts of the discussions.

The motions have a simple characterization in the Minkowskian frame
$\overline{\mathcal{R}}$. They correspond to inertial motions with a
uniform velocity $\eta c$ for light and $\overline{V}$ for matter
with $\left\vert \overline{V}\right\vert \ll c$. They may be labeled
by the set of associated conserved quantities, namely the energy
$\overline{E}$, the momentum $\overline{P}$ and the boost generator
\begin{equation}
\overline{K} =\overline{E}\overline{z}-\overline{P}c^{2}\overline{t}
=\overline{E}\overline{z_0}-\overline{P}c^{2}\overline{t_0} \,.
\label{atomsBar}
\end{equation}%
As $\overline{E}, \overline{P}, \overline{K}$ are conserved,
eq.\eqref{atomsBar} gives the equation of motion in
$\overline{\mathcal{R}}$ as well as in $\mathcal{R}$ (using
\eqref{CoorTra})
\begin{equation}
\overline{E}+\gamma\overline{K}=\left( 1+\gamma z\right) \left(
\overline{E} \cosh (\gamma ct) - \overline{P}c \sinh(\gamma
ct)\right) \,. \label{atomsLab}
\end{equation}%
Equation \eqref{atomsLab} also expresses conservation of the energy
$E$ defined in the stationary frame $\mathcal{R}$
\begin{equation}
E=\overline{E}\frac{\partial \overline{t}}{\partial
t}-\overline{P}\frac{\partial \overline{z}}{\partial
t}=\overline{E}+\gamma\overline{K} \,. \label{atomsE}
\end{equation}

The last equations may be written similarly for the motions of light
waves labeled by conserved quantities $\overline{e}, \overline{p},
\overline{k}$ (lowercase letters are used for light, uppercase
letters for atoms). The equation of motion is written
\begin{equation}
\overline{k}=\overline{e}\overline{z}-\overline{p}
c^{2}\overline{t}=-\eta \overline{e}\overline{u}^\eta
\label{lightBar}
\end{equation}%
for light waves with velocity $\eta c$ and momentum $\overline{p} =
\eta\overline{e}/c$ (assuming $x,y$ constant). It corresponds to a
constant light cone variable $\overline{u}^\eta$ and may be written
also in $\mathcal{R}$
\begin{eqnarray}
e=\overline{e}+\gamma\overline{k} = \overline{e}\left( 1+\gamma z
\right) e^{-\eta\gamma ct} \,. \label{lightLab}
\end{eqnarray}
This fixes relations between the quantities $e$, $\overline{e}$,
$\overline{k}$ and $\overline{u}^\eta$ which are conserved along
each light motion
\begin{equation}
\overline{e} = \frac{e} {1-\eta \gamma\overline{u}^\eta} \quad
,\quad \overline{k} = - \frac{\eta e\overline{u}^\eta}{1-\eta
\gamma\overline{u}^\eta} \,. \label{ebarkbar}
\end{equation}
This is also equivalent to the dependence of the Doppler effect on
position in an accelerated frame \cite{Einstein07,Jaekel96}.

Matter and light waves are coupled through stimulated Raman
interactions \cite{Kasevich92} at the vertices. Laser beams are
delivered by sources at rest at position $z=z^\mathrm{a}$ in
$\mathcal{R}$, which is translated into an accelerated motion in
$\overline{\mathcal{R}}$ \cite{Rindler77}
\begin{eqnarray}
\left( 1+\gamma z^{\mathrm{a}}\right) ^2 \nonumber &=&\left(
1-\gamma\overline{u}^+ \right) \left( 1 + \gamma \overline{u}^-
\right) \\ &=&\left( 1-\gamma\overline{z} \right) ^2-\left( \gamma
c\overline{t} \right) ^2 \,. \label{laserBar}
\end{eqnarray}%
The phases of these laser beams are discussed later on.

The geometry of the atomic gravimeter is sketched on the space-time
diagram of Figure 1. This diagram is drawn in the free fall frame
$\overline{\mathcal{R}}$ but it may be drawn as well in any other
frame, say $\mathcal{R}$. Differences between the drawings are only
{apparent} since they result from an arbitrary choice of the
coordinate system. Intrinsic statements bear on gauge-invariant
properties which do not depend on this choice. One aim of the
present paper is to associate gauge invariant evaluations and
statements to the geometrical drawing of Figure 1.

The conventions on Figure 1 have been chosen close to that of atomic
physics calculations \cite{Storey94}~: the space component $z$ is
along the ordinate axis and the time component $ct$ along the
abscissa; the units are chosen unequal in order to represent atomic
velocities (much smaller than $c$); the proportions are altered so
that light velocities are also apparent on the drawing. The
representation could be brought back to the one commonly used in
relativity (with $z$ along the abscissa and $ct$ along the ordinate
axis; see for example Fig.~1 in \cite{Dimopoulos08}) through a
mirror reflection with respect to the bisecting line between the two
coordinate axis.

\begin{figure}[t]
\centerline{\includegraphics[width=9cm]{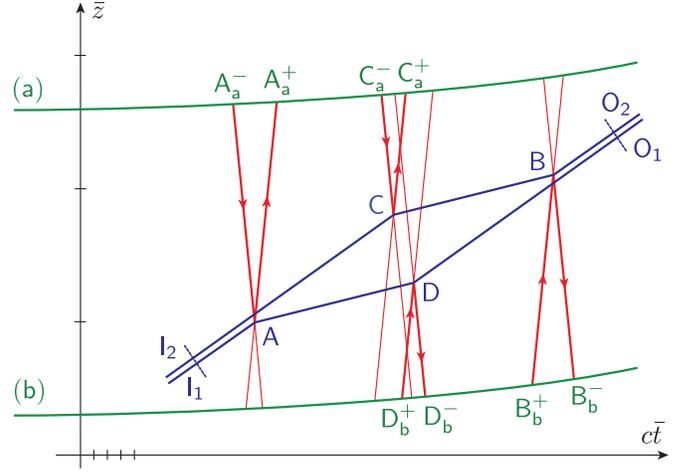}}
\caption{Space-time diagram for the atomic gravimeter in the free
fall frame $\overline{\mathcal{R}}$. Blue, red and green lines
represent the motions of atoms, light and laser sources
respectively. Stimulated Raman interactions take place at the
vertices $\A$, $\B$, $\C$, $\D$ where atomic and light lines
intersect. Light phases are delivered at the intersections of light
and laser lines (details in the text). The dotted lines $\I_1:\I_2$
and $\O_1:\O_2$ represent phase reference planes respectively at the
input and output of the interferometer. The ticks on the axes show
the units chosen for abscissa $c\overline{t}$ and ordinate
$\overline{z}$. [Colors online]} \label{Fig1}
\end{figure}

The blue, red and green lines on Figure 1 represent the motions of
atoms, light and lasers respectively (colors online). Two atomic
(blue) paths are represented, and the interference between these two
paths defines the dephasing (details below). On each path, inertial
segments are joined by stimulated Raman vertices $\A$, $\B$, $\C$,
$\D$ (letters chosen to match Fig.~9 in \cite{Storey94}). The photon
paths represent laser beams which have a propagation direction
indicated by a $\pm$ sign and cross atomic lines at vertices $\A$,
$\B$, $\C$, $\D$. The photon paths also intersect the laser lines
$(\a)$ and $(\b)$ which sit respectively {above} and {below} the
atoms. The intersection events are denoted by $\A^\eta_\ab$,
$\B^\eta_\ab$, $\C^\eta_\ab$ and $\D^\eta_\ab$ on Figure 1. The
drawing on Figure 1 is an idealized representation of the atomic
gravimeter with the experimental setup simplified as much as
possible while keeping the essential ingredients.

One of these essential ingredients is the link between phases and
conservation laws. As the atoms are freely falling in the time
intervals between interactions, they are accelerated with respect to
experimental platforms at rest on ground. The Doppler shift induced
by the free fall has thus to be compensated so that the stimulated
Raman transitions are on resonance at the different vertices (see
for example the footnote 49 in \cite{Dimopoulos08}). Two techniques
which have been used for this compensation \cite{Peters01,Cheinet06}
will be discussed later on and considered as illustrations of the
general expressions obtained in the present paper. There exists a
small tolerance for the compensation of the Doppler shift, which is
much more easily met under zero gravity conditions
\cite{vanZoest10,Geiger11}. In the present paper, we restrict our
attention to experiments performed in the typical Earth gravity
$g\sim9.81$m/s$^2$. Compensation of the Doppler shift induced by the
free fall is thus mandatory, and we have to account carefully for
the consequences of this compensation (details below).

In order to accommodate the various techniques used for that
compensation, we will consider the laser phases as functions
$\overline{\phi}^\eta\left(\overline{u}^\eta\right)$ of the
light-cone variables $\overline{u}^\eta$. These laser phases are
scalar quantities and may be written in any coordinate system
\begin{eqnarray}
\overline{\phi}^\eta\left(\overline{u}^\eta\right) =
{\phi}^\eta\left(t,z\right) \,. \label{light_pha}
\end{eqnarray}
The derivatives of these functions are the photon energies which are
involved in resonance conditions at vertices
\begin{eqnarray}
\overline{e}^\eta \equiv \hbar c \overline{\kappa}^\eta \quad,\quad
\overline{\kappa}^\eta \equiv \frac{\dd \overline{\phi}^\eta}{\dd
\overline{u}^\eta} \label{light_en} \quad, \\
{e}^\eta \equiv \hbar c {\kappa}^\eta \quad,\quad {\kappa}^\eta
\equiv \frac{\dd {\phi}^\eta}{c\dd t} \,. \nonumber
\end{eqnarray}
The transformation between $\overline{e}^\eta$ and ${e}^\eta$ is
given in \eqref{lightLab}.

The variation of the phase functions $\overline{\phi}^\eta$ fixes
the energies and, then, the shape of the interferometer through the
conservation laws. The same functions appear in the phase transfer
from light to matter at the vertices of the interferometer. We will
show in the next sections that these functions determine in fact all
the terms involved in the expression of the dephasing.

\section{Phases and conservation laws}

We now discuss phases for freely propagating atomic and light fields
as well as at interaction vertices. These elementary building blocks
will then be used to write the expression of the dephasing in the
interferometer.

Spin properties are disregarded throughout this paper, for matter
and light waves (see \cite{Borde89,Storey94} for discussions). Then,
the internal energies (energies in the atomic rest frame) are fixed
by atomic structure properties. They are the same on opposite atomic
lines of the interferometer (see Fig.~9 in \cite{Storey94}) with the
mass difference $M _\ast-M$ between adjacent lines defined by the
hyperfine structure ($\sim$9.2GHz for Caesium-133 atoms for
instance). Its value is small with respect to laser frequencies but
it is certainly not zero and has to be accounted for in relativistic
calculations \cite{Borde08}.

With each atomic line drawn on Figure 1 is associated a Klein-Gordon
field $\Psi(x)=F(x)e^{-i\Phi (x)}$ written as the product of a
slowly varying envelop $F$ and of a rapidly varying phase $\Phi$.
Similar relations hold for the light fields with lowercase letters
$\psi$, $f$, $\phi$ and $p_\mu$ replacing $\Psi$, $F$, $\Phi$ and
$P_\mu$. With each vertex $\A$, $\B$, $\C$, $\D$ on Figure 1 is
associated an equation relating the fields involved in the
stimulated Raman interaction. In a simplified approach, the outgoing
matter fields are given by incoming matter fields and light fields
(more general solutions discussed for example in \cite{Borde02})
\begin{eqnarray}
\Psi _\A^\out = \Psi _\A^\inp \psi _\A^- \left( \psi _\A ^+ \right)
^\ast \,,\, \Psi _\C^\out = \Psi _\C^\inp \psi _\C^- \left(
\psi _\C ^+ \right) ^\ast \,,&&  \nonumber \\
\Psi _\B^\out = \Psi _\B^\inp \psi _\B^+ \left( \psi _\B ^- \right)
^\ast \,,\, \Psi _\D^\out = \Psi _\D^\inp \psi _\D^+ \left( \psi _\D
^- \right) ^\ast \,.&& \label{vertex_wav}
\end{eqnarray}
The labels ``in'' and ``out'' correspond to incoming and outgoing
matter waves defined in the close vicinity of the vertex while ``-''
and ``+'' correspond to light waves propagating respectively
downward and upward. The light fields $\psi$ and $\psi^\ast$
correspond to positive and negative frequency parts of the waves and
they are associated respectively with absorption and stimulated
emission processes. Equations are the same at $\A$ and $\C$ where
the atom absorbs a downward photon and emits an upward one, and
changed through the exchange of +/- sign at vertices $\B$ and $\D$
where the atom absorbs an upward photon and emits a downward one.
This difference is indicated on the geometrical drawing of Figure 1
by thick red lines which bear the absorbed and emitted photons and
arrows which indicate the propagation directions. The multiplicative
relations \eqref{vertex_wav} for the waves are translated into
additive ones for the phases
\begin{eqnarray}
\Phi _\A^\out = \Phi _\A^\inp + \phi _\A^- - \phi _\A ^+ &,& \Phi
_\C^\out = \Phi _\C^\inp + \phi _\C^- - \phi _\C ^+ \,,\nonumber \\
\Phi _\B^\out = \Phi _\B^\inp + \phi _\B^+ - \phi _\B ^- &,& \Phi
_\D^\out = \Phi _\D^\inp + \phi _\D^+ - \phi _\D ^-\,.
\label{vertex_pha}
\end{eqnarray}

As energy and momentum are gradients of phases, the last equations
lead through differentiation to energy-momentum conservation laws
for the Raman interaction processes. We introduce lightcone
notations for energy-momentum variables in $\overline{\mathcal{R}}$
\begin{eqnarray}
&& \overline{P} ^\eta \equiv \frac12 \left( \frac{\overline{E}}c +
\eta \overline{P} \right) \label{lightcone_en}
\end{eqnarray}
in order to write the conservation laws
\begin{eqnarray}
\overline{P} _\A^{\eta,\out} = \overline{P} _\A^{\eta,\inp} -
\frac{\eta \overline{e} _\A ^\eta}c \,,\, \overline{P}
_\C^{\eta,\out} = \overline{P} _\C^{\eta,\inp} - \frac{\eta
\overline{e} _\C ^\eta}c \,, &&
\nonumber \\
\overline{P} _\B^{\eta,\out} = \overline{P} _\B^{\eta,\inp} +
\frac{\eta \overline{e} _\B^\eta}c \,,\, \overline{P}
_\D^{\eta,\out} = \overline{P} _\D^{\eta,\inp} + \frac{\eta
\overline{e} _\D^\eta}c \,. && \label{vertex_mom}
\end{eqnarray}
As the energies of the absorbed and emitted photons are close to
each other, the effective energy transfer is small while the
momentum transfer is significant, so that the main effect for the
atom is to undergo a kick due to momentum conservation.

As the atomic momenta are conserved on the inertial motions between
interactions, we can use the following relations to simplify
\eqref{vertex_mom}
\begin{eqnarray}
\overline{P} _\A^{\eta,\out} = \overline{P} _\D^{\eta,\inp} \equiv
\overline{P} _\AD^\eta &,& \overline{P} _\C^{\eta,\out} =
\overline{P} _\B^{\eta,\inp} \equiv \overline{P} _\CB^\eta \,.
\label{inert_mom}
\end{eqnarray}
We also assume that the interferometer is adjusted for plane waves
at its input and output ports. This means that momenta are identical
in the two incoming legs of the interfering paths as well as in the
outgoing legs
\begin{eqnarray}
\overline{P} _\A^{\eta,\inp} = \overline{P} _\C^{\eta,\inp} \equiv
\overline{P} _\I^\eta &,& \overline{P} _\D^{\eta,\out} =
\overline{P} _\B^{\eta,\out} \equiv \overline{P} _\O^\eta \,.
\end{eqnarray}
Using these relations, we rewrite the conservation laws
\begin{eqnarray}
\overline{P} _\AD^\eta = \overline{P} _\I^\eta - \frac{\eta
\overline{e} _\A ^\eta}c = \overline{P} _\O^\eta - \frac{\eta
\overline{e} _\D^\eta}c \,,
\nonumber \\
\overline{P} _\CB^\eta = \overline{P} _\I^\eta - \frac{\eta
\overline{e} _\C ^\eta}c = \overline{P} _\O^\eta - \frac{\eta
\overline{e} _\B^\eta}c  \,, \label{vertex_mom_}
\end{eqnarray}
and deduce
\begin{eqnarray}
\overline{e} _\A^\eta + \overline{e} _\B^\eta &=& \overline{e} _\C
^\eta + \overline{e} _\D^\eta \label{ident_freq} \\
&=& \eta c \left(\overline{P} _\I^\eta + \overline{P} _\O^\eta -
\overline{P} _\AD^\eta - \overline{P} _\CB^\eta \right) \,.
\nonumber
\end{eqnarray}

In order to account for all kinematical constraints, we also write
the equality of masses $M^2\equiv\overline{P} ^+\overline{P}
^-/4c^2$ on opposite atomic lines of the interferometer
\begin{eqnarray}
{M} _\I^2 = {M} _\O^2 \equiv M^2 &,& {M} _\AD^2 ={M} _\CB^2 \equiv
{M} _\ast^2 \,. \label{Masses}
\end{eqnarray}
The associated constraints are conveniently written by introducing
notations for the average values and differences of momenta on
opposite sides
\begin{eqnarray}
\overline{P}_\O^\eta \equiv \overline{P}^\eta + \frac{\Delta\overline{P}^\eta}2
&,& \overline{P}_\I^\eta \equiv \overline{P}^\eta - \frac{\Delta
\overline{P}^\eta}2 \,, \label{symmMom} \\
\overline{P}_\AD^\eta \equiv  \overline{P} _\ast^\eta + \frac{\Delta
\overline{P} _\ast^\eta}2 &,& \overline{P}_\CB^\eta \equiv
\overline{P} _\ast^\eta - \frac{\Delta \overline{P} _\ast^\eta}2 \,,
\nonumber
\end{eqnarray}
and deducing from \eqref{Masses}
\begin{eqnarray}
&&{\overline{e}_\D^\eta-\overline{e}_\A^\eta} =
{\overline{e}_\B^\eta-\overline{e}_\C^\eta} =
\frac{\overline{E}+\eta c\overline{P}}{2\overline{E}} {c\Delta
\overline{P}} \equiv {2\overline{\epsilon}^\eta} \,,
\label{solEn} \\
&&{\overline{e}_\C^\eta-\overline{e}_\A^\eta} =
{\overline{e}_\B^\eta-\overline{e}_\D^\eta} =
\frac{\overline{E}_\ast+\eta c\overline{P}_\ast}
{2\overline{E}_\ast} c\Delta \overline{P}_\ast \equiv
{2\overline{\epsilon}_\ast^\eta} \,. \nonumber
\end{eqnarray}

It follows that all photon energies can be expressed from the four
atomic momenta
\begin{eqnarray}
&&\overline{e}_\A^\eta = \overline{e}^\eta -
\overline{\epsilon}^\eta - \overline{\epsilon}_\ast^\eta \;,\;
\overline{e}_\B^\eta = \overline{e}^\eta + \overline{\epsilon}^\eta
+ \overline{\epsilon}_\ast^\eta\;,
\nonumber\\
&&\overline{e}_\C^\eta = \overline{e}^\eta -
\overline{\epsilon}^\eta + \overline{\epsilon}_\ast^\eta \;,\;
\overline{e}_\D^\eta = \overline{e}^\eta + \overline{\epsilon}^\eta
- \overline{\epsilon}_\ast^\eta\;, \label{photon_en}
\end{eqnarray}
where $\overline{e}^\eta$ denotes the mean photon energy (see
eqs.\ref{ident_freq})
\begin{eqnarray}
&& \overline{e}^\eta \equiv
\frac{\overline{e}_\A^\eta+\overline{e}_\B^\eta}2
=\frac{\overline{e}_\C^\eta+\overline{e}_\D^\eta}2 = \eta c
\left(\overline{P} ^\eta - \overline{P} _\ast^\eta  \right) \,.
\label{photon_en_mean}
\end{eqnarray}
As discussed in the previous section, the Doppler shift induced by
the free fall of the atoms must be compensated so that the
stimulated Raman transitions remain at resonance at the different
vertices. Equations (\ref{photon_en}-\ref{photon_en_mean}) provide
the relations to be fulfilled by the laser frequencies so that this
aim is met.

\section{Dephasing of the interferometer}

We now write the evolution of phases for free matter motions. We then
combine the obtained expressions with those of the previous section
to deduce the dephasing measured at the output of the interferometer.

For matter waves, the phase on an inertial motion, say the segment
$\AD$ on Figure 1, may be written under any of the equivalent forms
($\Phi_\AD \equiv \Phi_\D^\inp - \Phi_\A^\out$, $\tau_\AD \equiv
\tau_\D^\inp - \tau_\A^\out$ and $\overline{u}_\AD^\eta \equiv
\overline{u}_\D^\eta - \overline{u}_\A^\eta$)
\begin{eqnarray}
\hbar\Phi _\AD = {M_\AD c^2 \,\tau_\AD} = {2 \overline{P} _\AD^+
\overline{u}_\AD^+ } = {2 \overline{P} _\AD^- \overline{u}_\AD^- }
\,. \label{evol_pha}
\end{eqnarray}
The equivalence of these forms follows from the geometric identities
valid on any inertial motion
\begin{eqnarray}
\frac{2\overline{P}^+}{Mc} = \frac{Mc}{2\overline{P}^-} =
\sqrt{\frac{\overline{P}^+}{\overline{P}^-} } =
\sqrt{\frac{\dd\overline{u}^-}{\dd\overline{u}^+} } =
\frac{c\dd\tau}{\dd\overline{u}^+} =
\frac{\dd\overline{u}^-}{c\dd\tau} \,. && \label{identities}
\end{eqnarray}
Note also that the identity of the two last expressions in
\eqref{evol_pha} can be considered as a consequence of the
conservation of boost generator \eqref{atomsBar} along an inertial
motion
\begin{equation}
\overline{K} = \overline{P}^- \overline{u}^- - \overline{P}^+
\overline{u}^+  \quad,\quad \dd\overline{K} = 0 \,. \label{Boost}
\end{equation}

Under its differential form, \eqref{evol_pha} means that
$\hbar\dd\Phi$ may be written as proportional to elapsed proper time
$\dd\tau$ or lightcone variables increments $\dd\overline{u}^\eta$
\begin{eqnarray}
\hbar \dd\Phi = M c^2 \,\dd\tau = 2 \overline{P}^+ \dd\overline{u}^+
= 2 \overline{P}^- \dd\overline{u}^- \,. \label{evol_pha_int}
\end{eqnarray}
The first relation connects the methods of quantum theory, where
classical trajectories extremize the action integrals, with metric
theories, where geodesics extremize the proper time intervals
\cite{Landau}. The other relations imply that the matter phases are
functionals of the conserved quantities which characterize the light
rays involved at the interaction vertices. Note that for light
waves, the phase is constant ($\dd\phi =0$) or equivalently the
proper time is freezed ($\dd\tau=0$) along propagation. This is the
basic idea of Einstein synchronization and localization which allow
remote observers to exchange information by sharing the common value
of a light phase \cite{Jaekel96}. Here we see that the same light
phases are transcribed to atomic phases at the interaction vertices
and become an essential part of the dephasing measured by the
interferometer.

As the interferometer is adjusted for coherent matter waves at its
input and output ports, the dephasing $\Phi$ is just the difference
of phases accumulated on the two arms 1 and 2 (see Figure 1)
\begin{equation}
\Phi = \left( \Phi _{\O_2} - \Phi _{\I_2} \right) - \left( \Phi
_{\O_1} - \Phi _{\I_1} \right) \,. \label{def_dephasing}
\end{equation}
We collect the elementary relations written up to now, and deduce
the dephasing \eqref{def_dephasing} as follows
\begin{eqnarray}
\Phi &=& \Phi _\li + \Phi _\at \quad, \label{Phi_interf} \\
\Phi _\li &=& \Phi _\li^+ - \Phi _\li^- \quad,\quad \Phi _\li^\eta =
\phi _\A^\eta + \phi _\B^\eta - \phi _\C^\eta - \phi _\D^\eta \quad,
\nonumber\\
\Phi _\at &=& \left( \Phi _{\I_2\C} + \Phi _\CB + \Phi _{\B\O_2}
\right) - \left( \Phi _{\I_1\A} + \Phi _\AD + \Phi _{\D\O_1} \right)
\,. \nonumber
\end{eqnarray}%
$\Phi _\li$ is the sum of phases transferred from light to atoms at
the interaction vertices (see eqs.\ref{vertex_pha}) while $\Phi
_\at$ is the sum of propagation phases \eqref{evol_pha} over atomic
lines. Note that equation \eqref{Phi_interf} is independent on the
choice of the input and output reference planes $\I$ and $\O$ as
well as on the choice of points $\I_1$ and $\I_2$ in the plane $\I$
or, equivalently, of points $\O_1$ and $\O_2$ in the plane $\O$.
This is important since plane waves consist in coherent
superpositions of all points on these phase planes. It is also worth
stressing here that all terms appearing in \eqref{Phi_interf} have a
definition independent of the specific coordinate system used in the
calculation. This means that they are gauge invariant in the general
relativistic sense. One may remark that the sum $\Phi$ of light and
atomic terms is also invariant under changes of electromagnetic
gauge while being the physical observable measured in the
interferometer.

The expression of $\Phi _\at$ in \eqref{Phi_interf} contains the
\emph{separation terms} coming from the fact that the interferometer
has not necessarily a closed geometry \cite{Borde08,Dimopoulos08}. A
convenient convention is to have these reference planes just before
the first vertex and just after the last vertex so that
$\I_1\equiv\A\equiv\A_1$ and $\O_2\equiv\B\equiv\B_2$. Denoting
respectively $\A_2$ and $\B_1$ the points on trajectories 2 and 1
which lie on the same reference planes as $\A$ and $\B$, one deduces
\begin{eqnarray}
\Phi _\at &=& \left( \Phi _{\A_2\C} + \Phi _{\C\B_2} \right) -
\left( \Phi _{\A_1\D} + \Phi _{\D\B_1} \right) \,.
\end{eqnarray}%
Using the conservation laws \eqref{vertex_mom_}, we deduce that
$\Phi_\at$ may be written under either of the equivalent forms
\begin{eqnarray}
\Phi _\at &=& \Phi _\at^+ = - \Phi _\at^- = \frac{\Phi _\at^+ - \Phi
_\at^-}2 \,, \label{Phi_at}
\end{eqnarray}
where
\begin{eqnarray}
&&\Phi _\at^\eta = - 2 \left( \overline{\kappa} _\A^\eta
\overline{u} _\A^\eta + \overline{\kappa} _\B^\eta \overline{u}
_\B^\eta - \overline{\kappa} _\C^\eta \overline{u} _\C^\eta -
\overline{\kappa} _\D^\eta \overline{u} _\D^\eta \right)
+ \Phi _\sep^\eta \,, \nonumber\\
&&\Phi _\sep^\eta = 2\eta \frac{\overline{P}^\eta_\O \left(
\overline{u}^\eta_{\B_2} - \overline{u}^\eta_{\B_1} \right) -
\overline{P}^\eta_\I \left( \overline{u}^\eta_{\A_2} -
\overline{u}^\eta_{\A_1} \right) }\hbar \,. \label{Phi_at_}
\end{eqnarray}
In \eqref{Phi_at_}, the separation terms $\Phi _\sep^\eta$ are
written in terms of the input and output atomic momenta whereas the
other terms in the first line depend only on the momenta transferred
at vertices.

In order to make the discussion more explicit, let us now consider
the case of a closed geometry with the interferometer adjusted so
that $\B_1$ and $\B_2$ coincide when $\A_1$ and $\A_2$ coincide.
After this adjustment, the separation terms $\Phi _\sep^\eta$ vanish
and the dephasing is given by a compact expression collecting all
contributions
\begin{eqnarray}
\Phi &=& \Phi ^+ - \Phi ^-  \quad, \label{Phi_interf_closed} \\
\Phi ^\eta &=& \varphi _\A^\eta + \varphi _\B^\eta - \varphi
_\C^\eta - \varphi _\D^\eta \quad, \nonumber\\
\varphi^\eta &\equiv& \phi^\eta - \overline{\kappa}^\eta
\overline{u}^\eta = \phi^\eta - \overline{u}^\eta \frac{\dd
{\phi}^\eta}{\dd \overline{u}^\eta} \,. \nonumber
\end{eqnarray}
Note that $\varphi^\eta$ is the Legendre transform of the laser
phase $\phi^\eta$ and is more naturally considered as a function of
the laser wavevector $\overline{\kappa}^\eta$ with the relations
$\dd {\phi}^\eta=\overline{\kappa}^\eta\dd \overline{u}^\eta$ and
$\dd {\varphi}^\eta=-\overline{u}^\eta\dd \overline{\kappa}^\eta$.

The forthcoming discussions will be presented in this case of a
closed geometry where the dephasing is given by the expression
\eqref{Phi_interf_closed}, completely characterized by the Legendre
transform of the laser phases at the 4 events $\A$, $\B$, $\C$ and
$\D$. Note that, if needed, one may come back to the general case by
adding the separation terms written in \eqref{Phi_at_}.

\section{Discussion}

We now illustrate the expression \eqref{Phi_interf_closed} obtained
in the preceding section for a closed interferometer by discussing
two techniques which have been used for compensating the Doppler
shift. One such technique corresponds to chirped frequencies with
the chirp parameter being adjusted \cite{Cheinet06}. Another one,
described by Figure.~13 in \cite{Peters01}, corresponds to ramped
variations, with Raman laser frequencies constant by pieces and
changed between vertices.

We first consider the case where the laser frequencies are chirped
so that resonance is preserved with freely falling atoms
\cite{Cheinet06}. This can be described within the following family
of parameterizations of light phases
\begin{equation}
\phi ^\eta = \phi _0^\eta - \frac{\omega^\eta}c \left(
\frac{1-\alpha}{\eta\gamma} \ln\left( 1-\eta \gamma
\overline{u}^\eta \right) - \alpha \overline{u}^\eta \right) \,.
\label{phi_chirp}
\end{equation}
The variations of associated energies are deduced from
\eqref{light_en} (with $\omega^\eta$, $\alpha$ and $\phi _0^\eta$
constant)
\begin{eqnarray}
&&\overline{e}^\eta = \hbar c \overline{\kappa}^\eta =
\hbar\omega^\eta \frac{1-\eta \alpha \gamma \overline{u}^\eta
}{1-\eta \gamma \overline{u}^\eta } \quad, \nonumber \\
&&{e}^\eta = \hbar c \kappa^\eta = \hbar\omega^\eta \left(1-\eta
\alpha \gamma \overline{u}^\eta \right) \,. \label{chirp}
\end{eqnarray}
With the chirp parameter $\alpha$ set to 0, ${\kappa}^\eta$ is
constant and the interferometer cannot operate properly because of
the Doppler shift induced by free fall. With $\alpha$ set to 1 in
contrast, $\overline{\kappa}^\eta$ is constant, and the resonance
conditions are met at the 4 vertices in a symmetric geometry where
the interferometer has the shape of a parallelogram. Atomic momenta
are equal on opposite sides, so that atoms spend equal proper times
on opposite sides. It follows that the atomic part of the dephasing
is obviously zero ($\Phi _\at=0$) as a consequence of this symmetry.

This symmetry also entails that the light part of the dephasing
vanishes, though in a more indirect manner. As a matter of fact, the
momenta $\overline{\kappa}^\eta$ are equal at the 4 vertices (for
each value of $\eta$ taken separately), so that the derivatives of
the phase functions $\phi^\eta$ are equal at $\A$, $\B$, $\C$ and
$\D$. Within the family of parameterizations \eqref{phi_chirp}, the
derivatives $\overline{\kappa}^\eta$ are thus constant everywhere
(not only at the vertices), and phases $\phi^\eta$ are linear
functions of the lightcone variables $\overline{u}^\eta$. Then the
light part $\Phi_\li$ of the dephasing also vanishes because of the
symmetry of the interferometer. This remark has been used as the
starting point for a measurement procedure for the acceleration.
This is the procedure explained in \cite{Cheinet06}~: when the chirp
parameter $\alpha\gamma$ appearing in the law of variation
\eqref{chirp} of $\kappa^\eta$ is varied in parameterizations
\eqref{phi_chirp}, both contributions $\Phi_\at$ and $\Phi_\li$ to
the dephasing $\Phi$ vanish simultaneously at $\alpha=1$. The
specific value of this chirp parameter producing a null dephasing
thus leads to a determination of the value of the acceleration
parameter $\gamma$.

Equation \eqref{Phi_interf_closed} can be used for a fast
demonstration of the fact that the full dephasing vanishes in this
case. As a matter of fact, when $\alpha=1$ in \eqref{phi_chirp}, the
Legendre transform $\varphi^\eta$ of the phase is constant, so that
the combination $\varphi _\A^\eta + \varphi _\B^\eta - \varphi
_\C^\eta - \varphi _\D^\eta$ appearing in \eqref{Phi_interf_closed}
is obviously zero. This fast line is also useful for evaluating the
dominant terms in the vicinity of $\alpha=1$ (development in the
small parameter $\alpha-1$)
\begin{eqnarray}
\frac{c \Phi ^\eta}{\omega^\eta} &=& \left(\alpha-1\right)
\frac{\eta\gamma}2 \left(\overline{u}^{\eta\ 2}_\A +
\overline{u}^{\eta\ 2}_\B - \overline{u}^{\eta\ 2} _\C -
\overline{u}^{\eta\ 2} _\D \right)
\label{Phi_interf_closed_chirp} \\
&+& \left(\alpha-1\right) \frac{2\gamma^2}3
\left(\overline{u}^{\eta\ 3}_\A + \overline{u}^{\eta\ 3}_\B -
\overline{u}^{\eta\ 3} _\C - \overline{u}^{\eta\ 3} _\D \right)+
\ldots \nonumber
\end{eqnarray}

As the higher order terms decrease rapidly since
$\gamma\overline{u}$ are small dimensionless numbers, this
expression is dominated by the second order term. In perturbation
theory, one may then calculate this term in the symmetric
configuration which corresponds to $\alpha=1$. In order to do so, we
introduce the middle point $\M$ of the parallelogram and use the
following relations
\begin{eqnarray}
&&\overline{u}^\eta_\B - \overline{u}^\eta_\M = \overline{u}^\eta_\M
- \overline{u}^\eta_\A = \frac{\overline{u}^\eta_\B -
\overline{u}^\eta_\A}2 \,,\nonumber \\
&&\overline{u}^\eta_\D - \overline{u}^\eta_\M = \overline{u}^\eta_\M
- \overline{u}^\eta_\C = \frac{\overline{u}^\eta_\D -
\overline{u}^\eta_\C}2 \,.
\end{eqnarray}
We then obtain an expression of the dephasing in terms of the
diagonals of the parallelogram
\begin{eqnarray}
\frac{c \Phi ^\eta}{\omega^\eta} \simeq \left(\alpha-1\right)
\frac{\eta\gamma}4 \left((\overline{u}^\eta_\B -
\overline{u}^\eta_\A)^2 - (\overline{u}^\eta_\D -
\overline{u}^\eta_\C)^2 \right) \,. &&
\end{eqnarray}
If one also uses the approximations that the atoms have non
relativistic velocities and that the momenta transferred at vertices
are small, one finally obtains
\begin{eqnarray}
\Phi &\sim& k_\eff\left(\alpha-1\right) \gamma c^2T^2 \quad,\quad
k_\eff= \frac{\omega^+ + \omega^-}c \,.
\label{Phi_interf_closed_chirp_approx}
\end{eqnarray}
As is well-known, the dephasing is thus determined by the laser
wavevectors entering the definition of $k_\eff$ (and not by the
Compton wavelength associated with the mass of the atoms).

We now discuss another measurement procedure, which corresponds to
the ramped phase functions sketched on Fig.~13 in \cite{Peters01}.
In this case, $\kappa$ is supposed to be constant during intervals
of time and to undergo jumps between the vertices so that resonance
conditions are still met (at least approximately) at the 4 vertices.
The phase functions thus correspond to \eqref{phi_chirp} with
$\alpha=0$ on three intervals $i=1,2,3$ containing $\A$ for the
first, $\C$ and $\D$ for the second, $\B$ for the third
\begin{equation}
\phi ^\eta = \phi _i^\eta - \frac{\omega^\eta_i }{\eta\gamma c}
\ln\left( 1-\eta \gamma \overline{u}^\eta \right) \,.
\label{phi_ramp}
\end{equation}
Accordingly, the energies correspond to \eqref{chirp} with
$\alpha=0$
\begin{eqnarray}
\overline{e}^\eta = \hbar c\overline{\kappa}^\eta =
\frac{\hbar\omega^\eta_i}{1-\eta  \gamma \overline{u}^\eta }
\quad,\quad {e}^\eta = \hbar c{\kappa}^\eta = \hbar\omega^\eta_i \,.
\label{ramp}
\end{eqnarray}
The energies are discontinuous at the jumps while the phases are
continuous \cite{Peters01}.

The parameters are then adjusted so that phases and energies at the
vertices are as close as possible to those obtained in the chirped
case. This can be done exactly at $\A$, $\B$, and $\M$ but not at
the two points $\C$ and $\D$. There remain small defects because
this compensation technique cannot meet the resonance prescription
at the two points $\C$ and $\D$. One deduces in particular from
\eqref{ramp} that
\begin{eqnarray}
\frac{\overline{\kappa}^\eta_\D}{\overline{\kappa}^\eta_\C} =
\frac{1-\eta \gamma \overline{u}^\eta _\C}{1-\eta \gamma
\overline{u}^\eta _\D} \,. &&
\end{eqnarray}
Meanwhile, this ratio should be 1 for the compensation to be exact,
which corresponds to the fact that the phases \eqref{phi_ramp} do
not reproduce exactly the law \eqref{phi_chirp} with $\alpha=1$
which led to a null dephasing. The order of magnitude of these
defects is however small as the two points $\C$ and $\D$ are close
to each other in the conditions of the experiments \cite{Peters01}.
This entails that this second method leads to results close to that
of the first one.

\section{Conclusion}

In the present paper, we have emphasized the status of relativistic
observable of the phase measured by an atomic gravimeter. In
particular, we have written it in terms of laser phases which are
invariant under relativistic gauge transformations (they do not
depend on the choice of the coordinate system). The physical
observable $\Phi$, which is the sum of light and atomic
contributions to the dephasing, is also invariant under
electromagnetic gauge transformations.

We have also studied in a detailed manner the interplay of the
expressions thus obtained with conservation laws at the interaction
vertices. The general expression of the dephasing is given by
equations \eqref{Phi_interf} and \eqref{Phi_at_} for an arbitrary
geometry and by the more compact form \eqref{Phi_interf_closed} when
the interferometer is adjusted to have a closed geometry. This
compact form, which is given by a Legendre transform of the laser
phases, makes obvious the intimate connection between light and
atomic contributions to the dephasing.

We have finally illustrated these general expressions by discussing
two techniques which have been used for compensating the Doppler
shift, one corresponding to chirped frequencies  and the other one
to ramped variations. We have thus recovered the known result that
the dephasing is determined by the momenta transferred at the
vertices, \textit{i.e.} by the laser wavevectors (and not by the
Compton wavelength associated with the mass of the atoms).

\acknowledgements{Thanks are due for fruitful discussions to L.
Blanchet, C.J. Bord{\'e}, P. Bouyer, C. Cohen-Tannoudji, D.M.
Greenberger, A. Lambrecht, A. Landragin, C. Salomon, W. Schleich and
P. Wolf.}

\end{document}